\documentclass[manuscript]{aastex}
\slugcomment{Submitted to Ap.J.}
\shortauthors{Dhawan, Mirabel, \& Rodr\'\i guez}
\shorttitle{AU-Scale Jets in GRS~1915+105}                 

\begin{document}

\title{AU-Scale Synchrotron Jets and Superluminal Ejecta in GRS~1915+105.}

\author{V. Dhawan}
\email{vdhawan@nrao.edu} 
\affil{National Radio Astronomy Observatory, Socorro, NM 87801 
\altaffilmark{1}}

\author{I. F. Mirabel}
\email{mirabel@discovery.saclay.cea.fr}
\affil{CEA/DSM/DAPNIA/SAp, Centre d'Etudes de
Saclay, F-91191 Gif-sur-Yvette, France \& 
Instituto de Astronom\'\i a y F\'\i sica del Espacio, Buenos Aires, Argentina}

\and

\author{L. F. Rodr\'\i guez}
\email{luisfr@astrosmo.unam.mx}
\affil{Instituto de Astronom\'\i a, UNAM, Apdo. Postal 70-264, 04510,
M\'exico, DF, M\'exico}

\altaffiltext{1}{The NRAO is a facility of the National Science
Foundation operated under cooperative agreement by Associated
Universities, Inc.\ }

\begin{abstract}

Radio imaging of the microquasar GRS~1915+105 with the Very Long
Baseline Array (VLBA) over a range of wavelengths 
(13, 3.6, 2.0 and 0.7~cm), in different states of the black hole
binary, always resolves
the nucleus as a compact jet of length $\sim$10$\lambda_{\rm cm}$~AU.
The nucleus is best imaged at the shorter wavelengths, on scales of
2.5~-~7~AU (0.2~-~0.6~mas resolution).  The brightness temperature of
the core is T$_{\rm B}\geq$10$^{9}$ K, 
and its properties are better fit by a conically expanding
synchrotron jet model, rather than a thermal jet.  The nuclear jet
varies in $\sim$30~min during minor X-ray/radio  outbursts, and
re-establishes within $\sim$18~hours of a major outburst, indicating
the robustness of the X-ray/radio (or disk/jet) system to disruption.

At lower resolution (80-240~AU), more extended ejecta are imaged at
$\sim$500 AU separation from the stationary core. 
Time-lapse images clearly detect the superluminal motion of the ejecta 
in a few  hours. The measured velocity is 1.5$\pm$0.1~c~(D/12~kpc) 
for the approaching
component, and is consistent with ballistic motion of the ejecta from
500~AU outwards, perhaps even since birth.  The axis of the
ejecta differs by $\le$12$^{\circ}$ clockwise from the
axis of the AU-scale jet, measured in the same observation. Both axes
are stable in time ($\pm$5$^{\circ}$), the AU scale for two  years, and
the large scale for over four years.
Astrometry over two years relative to an extragalactic reference
locates the black hole to $\pm$1.5~mas, and  its secular parallax due
to Galactic rotation is 5.8$\pm$1.5~mas~yr$^{-1}$, consistent with a
distance of 12~kpc. Finally, a limit of $\le$100~km~s$^{-1}$ is placed
on its proper-motion with respect to its neighbourhood.

Some accreting black holes of stellar mass (e.g. Cyg~X-1,
1E~1740-2942, GRS~1758-258, GX~339-4) and supermassive black holes
at the centre of galaxies (e.g. Sgr~A$^*$) lack evidence of large flares
and discrete transient ejecta, but have compact radio cores with
steady, flat-spectrum `plateau' states, like GRS~1915+105. Until now
GRS~1915+105 is the only system where both AU-scale
steady jets and large-scale superluminal ejections have been
unambiguously observed. Our observations suggest that the
unresolved flat-spectrum radio cores of accreting black
holes are compact quasi-continuous synchrotron jets.

\end{abstract}


\section{Introduction}

The X-ray transient GRS~1915+105, discovered in 1992 by the \sl Granat
\rm satellite, \citep{cas94}, is one of a few Galactic sources
exhibiting superluminal radio ejecta (see \citet{araa99} for a
review).  This microquasar offers a nearby laboratory for the study of
black hole accretion, and the associated phenomena of jet formation,
collimation, and outflow.  Since the characteristic evolution time is
proportional to the mass of the central black hole (Rees, 1998; Sams et
al. 1996), microquasars can reveal in minutes a richness of phenomena
analogous to a 10$^8$~M$_{\odot}$ quasar observed over centuries.

In this paper, we present radio images of the microquasar GRS~1915+105,
made with the Very Long Baseline Array (VLBA, see \citet{vlba95}), 
in two distinct states
of the black hole binary, the {\bf plateau} and {\bf flare} states. The
{\bf plateau} state is characterized by a flat radio spectrum, compact
size of a few AU, and flux density 10-100~mJy. The XTE (2-12~keV) soft
X-rays are weak, and the BATSE (20-100~keV) emission is strong. In
contrast, during the {\bf flare} state, with rise-time $\le$1~day,
optically thin ejecta of up to 1~Jy at $\lambda$13~cm
(S$_{\nu}~\propto \nu^{-0.6}$), are expelled to thousands of AU and fade
over several days. The soft X-rays also flare and show extreme
variability, while the hard X-rays fade for a few days before
recovering.  Observations over a range of wavelengths (13~cm, 3.6~cm,
2~cm and 0.7~cm), distinguish the intrinsically elongated nucleus, of
length $\sim$10$\lambda_{\rm cm}$~AU, from angular broadening due to
scattering by the line-of-sight ionized interstellar medium in the
Galactic plane.

GRS~1915+105 was also imaged in its distinct state of pronounced X-ray
dips, that repeat on timescales ranging from $\sim$12-60~min, with the
same repetition rate seen in the IR and radio pulses. The radio and
X-ray variability of GRS~1915+105 showed early evidence for recurrent
accretion and jet formation \citep{rod99,fos96}. There now exists a
wealth of data from X-ray, IR and radio observations, offering clues to
the coupled evolution of accretion disk, corona, and jet. On several
occasions, the flux of thermal X-rays from the inner accretion disk has
been observed to diminish, with simultaneous hardening of the power-law
spectrum from relativistic electrons, followed by progressively delayed
emission in the IR and radio,
\citep{mir98,mar99,bel97b,eik98,poo97,fen97,fen98}.  One scenario is to
have the inner part of the accretion disk and corona accelerated to
relativistic speeds, and ejected as synchrotron-emitting plasma.
Alternatively, advective infall of the disk past the black hole horizon
\citep{nar97}, could account for the X-ray dips, but not for the IR and
radio emission. A solution allowing simultaneous inflow and outflow 
is presented in the `ADIOS' (Advection Dominated Inflow-Outflow
Solution)  model of \citet{bla99},
but mechanisms to produce the observed X-ray variability without ADAF
(Advection Dominated Advection Flow) are discussed in \citet{nay00}.  
Though the mechanism is still debated, this particular source switches 
the jets on and off,
presenting repeated but unpredictable opportunities to observe the
ejecta evolve from  scales of a few to thousands of AU.

\section{VLBA Observations \label{sec-vlbaobs}}

Triggered by GBI \footnote{The Green Bank Interferometer is a facility
of the National Science Foundation operated by NRAO with support of the
NASA High Energy Astrophysics programs.}\ daily monitoring, we have
coordinated observations with X-ray  and IR telescopes, followed the
radio variations with the VLA \citep{mir98}, and imaged the radio
emission at multiple wavelengths with the VLBA (this paper).  Crucial
capabilities of the VLBA were the dynamic scheduling in
response to transients, and phase-referencing which permits astrometry
and the  detection of weak sources.

The VLBA observation epochs and parameters are presented in Table-1.
Observations are possible in any of nine wavelength bands, with band
changes requiring 20~s or less.  Simultaneous data can be obtained from
the 3.6~cm/13~cm receiver pair, by means of frequency-selective optics.
The choice of band was made just before each observation,  based on the
most recent flux and spectral information. Observations at 2~cm and
7~mm are least affected by galactic electron scattering, and were used to
observe the AU scales when the radio-to-IR spectrum was flat. During
the steep-spectrum flares, we employed the 3.6~cm/13~cm pair to image
large-scale ejecta. Other bands were observed for $\sim$20-40~mins in
rotation.  We employed the highest data recording rates, (256 or
128~Mb~s$^{-1}$, subject to scheduling constraints), for maximum
sensitivity in short time, since the source is known to be highly
variable. Ultimately, our image quality is limited not by thermal
noise, but by the restriction of instantaneous UV coverage to a maximum
of 45 baselines (Fourier components) from 10 antennas.

\placetable{table-1}

All the VLBA observations (except 7~mm) were phase-referenced, i.e.,
the antennas alternated between the target and one of several
extragalactic calibrators of known (sub-millarcsec) position. Cycle
times of three minutes were adequately fast to follow tropospheric
phase fluctuations at 2~cm and longer wavelengths. Thus the
interferometer array was rendered coherent for many hours, and the
resulting astrometric accuracy of $\sim$1.5~mas allows us to
discriminate between moving and fixed source components. In addition,
with frequent amplitude calibration, we avoid blind self-calibration on
the time-variable target.

We generated time-resolved images, with snap-shot intervals ranging
from 5 to 60~mins, depending on the beam size. Self-calibration was
used self-consistently, i.e. only over the snapshot interval.
Snap-shots were averaged in the image plane and UV data were combined
over longer intervals only if the snapshots were indistinguishable. In
practice, there is little effect at lower resolutions from the moving
ejecta - mainly a blurring of the moving component, and an increase in
residual artifacts, since self-calibration does not work perfectly if
the structure is varying. (See e.g., Fig.~\ref{fig-2}B, where the
extension of the SW component away from the core has occurred during
the `exposure time' of 5.1~hr.)

For the high-resolution images (e.g. in Fig.~\ref{fig-5}), the problem
of reliable imaging is more severe. In addition to snapshot imaging, a
careful search for time-variable or moving structure was done by
examining and model fitting to the closure phases over intervals
as short as 5~mins. Simultaneous flux measurement (from the VLA or GBI, when
available) was used to constrain the total flux while modelling the
AU-scale structure.  No evidence for rapidly moving or variable
structures was found. We can rule out discrete, moving ejecta on AU
scales - none of the snapshots or closure phases showed, e.g. double
sources or differently oriented structures.  We cannot at this time
rule out fast-moving features with low contrast, that would be blurred
out. The simulations can be used to place limits on the artifacts due
to the technique, to be discussed in a future paper.  We will continue
to pursue evidence for AU-scale motion in new data, as well as old data
after further refinement in software to model moving, time-variable
sources.

\section{Images During Two Flares}

These target-of-opportunity observations are subject to the constraints
of weather, scheduling, and source variability. For clarity, we present
individual images for two episodes of flare activity, and merge the
discussion of the common elements thereafter.

\subsection{1997 October}

The radio and X-ray behavior of GRS~1915+105 around this flare is
summarized in Fig.~\ref{fig-1}.  The AU-scale jet is imaged in the
nucleus (Fig~\ref{fig-2}A) on 1997 October 23, six days before a major
flare.  Images on October 31, centered at MJD~50752.02,
(Fig.~\ref{fig-2}B \& C) show the central core with a bright component
47.5$\pm$0.3~mas away on the SE (approaching) side.  The core has a
flux density of $\sim$20~mJy with flat spectrum, while the SE ejected
component has S$_{\nu} \propto ~\nu^{-0.5}$.
 The position of the SE component corresponds to a separation rate of
0.90$\pm$0.05~mas~hr$^{-1}$ in 53.3$\pm$2.4~hrs since the estimated
start of the flare at MJD=50749.8. A NW (receding) component is
marginally detected, at 17.4$\pm$0.3mas separation, consistent with it
being the counter-ejection, with velocity 0.33$\pm$0.02~mas~hr$^{-1}$
and flux density only $\sim$10\% of the SE component.

The time-lapse images, (Fig.~\ref{fig-3}), with interval of 2.5~hrs,
show a separation change of the SE ejection from the core of
2.3$\pm$0.2~mas, or 0.92$\pm$0.08 mas~hr$^{-1}$.  Our start time for
the flare is derived by extrapolating the slopes of the X-ray and radio
data (see Fig.~\ref{fig-1}). With this start time the velocity derived
from the position of the ejecta after 53.3~hours agrees with the
velocity from the 2.5~hr time lapse images.  \citet{fen99} assume a
considerably different start at 50750.5 for the same flare.  However,
their velocities for the ejecta, derived from images several days after
the flare, at 500 to 5000~AU scales, agree with ours within the errors,
and imply ballistic motion from a few to 1000's of AU, certainly beyond
500~AU.

\placefigure{fig-1}

\placefigure{fig-2}

\placefigure{fig-3}          

Astrometry (see Fig.~\ref{fig-7}) before and after the flare locates
the nucleus within 1.5~mas of the position we have measured for over
two years, after allowing for secular parallax. The 2~cm flux density
varies with the 30~min period seen in XTE dips, starting at 50750.5
\citep{fen99}. We show that the variable radio emission is from the
AU-scale jet, see  Fig.~\ref{fig-9}; (see also
\citet{mir98,poo97,fen97}).  We conclude that the nuclear jet that we
image has re-established itself within $\sim$18~hrs of the start of a
major outburst, if indeed it was disrupted at all.  The position angle
of the AU-scale jet in Fig.~\ref{fig-2}A is 157$\pm$2$^{\circ}$ at
2~cm, whereas the large-scale ejecta show 133$\pm$3$^{\circ}$ at 13~cm,
in Fig.~\ref{fig-2}B; and 143$\pm$4$^{\circ}$ at 3.6~cm in
Fig.~\ref{fig-2}C.

\subsection{1998 April-May}

The radio and X-ray behavior of GRS~1915+105 around this flare is
summarized in Fig.~\ref{fig-4}.  As in the 1997 October event, there is
a jet in the nucleus during the plateau state, (see Fig.~\ref{fig-5}D),
 two days before the flare which started on MJD$\sim$50916. Figures 5E
and 5F also show compact jets 2 days after the start of a second flare
on MJD$\sim$50932.  The spectrum of the jet is essentially flat from
2~cm to 0.7~cm.  After two intervening flares, on MJD~50935.5, the nucleus
is still within 1.0~mas of the expected position, after allowing for
secular parallax.  The position angles of the AU-scale jet are as
follows:  D:~155$\pm$2$^{\circ}$, at 2~cm; E:~154$\pm$4$^{\circ}$ again
at 2~cm; F:~145$\pm$6$^{\circ}$, at 0.7~cm for the overall structure,
although the innermost contours are rotated to 168$\pm$5$^{\circ}$,
perhaps indicating a bent jet.

The phase-referenced image at 2~cm (Fig.~\ref{fig-6}), heavily tapered
to 100~AU beam size, shows the SE ejection at $\sim$650~AU from the
core.  The counter-ejection was not detected, with flux $\le$5.5\% of
the approaching component.  The separation from the core of the SE
component corresponds to a motion of 57.5$\pm$0.5~mas in 67$\pm$7~hrs,
or 0.87$\pm$0.10~mas~hr$^{-1}$. This is consistent with the velocity of
0.93$\pm$0.07~mas~hr$^{-1}$, seen in the red and blue snapshots in
Fig.~\ref{fig-6}, which are 4.5~hrs apart.  The position angle of the
SE component is 148$\pm$4$^{\circ}$ from the core.

\placefigure{fig-4}

\placefigure{fig-5}

\placefigure{fig-6}              

\section{Astrometry \label{sec-astro}}

Astrometry is vital to register moving components from one epoch to the
next. The present astrometric accuracy is achieved with phase residuals
to the VLBA correlator model, and no special software.  The correlator
model used in these observations was in error by $\sim$30~mas for Earth
nutation, reduced by a factor of 0.1 due to the $\sim$5$^{\circ}$
distance to the calibrator.  We thus expect a maximum error in the
absolute position determination of 3~mas for a given epoch.  In fact,
when we compare our positions for secondary calibrators in these
observations to those obtained independently for the same calibrators
in the same reference frame by USNO, we find agreement within 1.5~mas,
(T.M. Eubanks, private communication).  The effects of nutation error
rate on proper motion are further reduced by the time-differencing, and
are $<$0.3~mas~yr$^{-1}$. Our errors in the secular parallax are
$\sim$1.5~mas~yr$^{-1}$, dominated by tropospheric `seeing', i.e.,
residual phase errors from the calibrator. This can be seen in
Fig.~\ref{fig-7} as the short-term scatter of positions over a few
hours on  a given date.

\placefigure{fig-7}

The J2000 position of the core is 19$^h$15$^m$11$^s$.54938$\pm$.00007,
10$^{\circ}$56'44''.7585$\pm$.001, on 1998 May 02,in the ICRF reference
frame \citep{ma98}. The secular parallax is -5.4$\pm$1~mas~yr$^{-1}$
and -2.3$\pm$1~mas~yr$^{-1}$ in RA and Dec respectively. This level of
astrometric accuracy, though it can be improved by further analysis, is
adequate to demonstrate that:

(i) The identification of the stationary core and moving ejecta is
unambiguous. We note that the 1997 October 31 data at 3.6~cm are shifted
from the pre-flare position by $\sim$2~mas, probably due to the complex and
evolving structure during the flare. However, this error is not enough
to cause mis-identification of core with ejecta separated by 47.5~mas.

(ii) GRS1915 shows the galatic rotation expected from an object about
12~kpc distant, \citep{dha00}, at the position
$l$~=~45$\rlap.{^\circ}$37, $b$~=~-0$\rlap.{^\circ}$22.  We find a
proper motion of 5.8$\pm$1.5~mas~yr$^{-1}$, which we ascribe to the
secular parallax of the core, within errors of $\pm$75~km~s$^{-1}$ in
the plane of the Galaxy (all errors are $\pm$3$\sigma$).  A
model-independent distance may perhaps be determined in the future by
measuring the annual trigonometric parallax of $\sim \pm$80
$\mu$arcsec, using closer calibrator sources, as attempted
 in the case of Sgr~A$^*$ \citep{reid99}.

(iii) The core is stationary on the sky to $\pm$1.5~mas once the
secular parallax is accounted for. The systematic drift perpendicular
to the Galactic plane over two years is consistent with
0$\pm$50~km~s$^{-1}$.

Combining errors parallel and perpendicular to the galactic plane, we
can place an upper limit of $<$100~km~s$^{-1}$ for the velocity of
GRS~1915+105 on the sky. Note that velocity along the line of sight is
unconstrained by these observations, but could become evident in the
Doppler shift of periodic features, if any are detected in the future.

\section{Scattering in the ISM Towards GRS~1915+105. \label{sec-scat}}

GRS~1915+105 lies near the galactic plane, along a line of sight
tangential to a spiral arm and intercepting a large column of gas, both
neutral and ionized. The high dispersion measure of distant pulsars in
this direction \citep{tay93} leads us to expect considerable angular
broadening due to the inhomogeneous ionized interstellar medium, as is
indeed the case.

Observation over several octaves can separate the intrinsic structure
from the scatter-broadening, as shown in Fig.~\ref{fig-8}, and short
wavelengths (2~cm and 0.7~cm) were used to give an unscattered view into
the nucleus where the jet originates. The minor axis size scales as
$\lambda^2$, and we measure a  scattering size of 1.9$\pm$0.1mas at 
3.6~cm, (135~mas at 30~cm=1~GHz).
Interestingly, the scattering seems to vary dramatically on scales of
$\sim$50~pc, seen by comparing our result with the 8.4~mas at 30~cm,
(measured by \citet{kem88}) for the OH maser in OH45.47+0.13, which
lies beyond the tangent point, as is the case for GRS~1915+105, and
only 10' away on the sky. A fraction of this increase by a factor of 16
in scattering  could be due to a putative
ionized circumstellar cocoon around the X-ray binary, as suggested by
recent infrared spectroscopy with the VLT, \citep{mart20}.

\placefigure{fig-8}

\section{Two Distinct Radio Emission Phenomena?}

We discuss in this section the evidence for two distinct radio emission
states, described most simply as attached to, and detached from the
nucleus.  They correspond to: \\
(a) A smooth, flat spectrum, continuous nuclear jet at AU scale, with
quiescent or oscillating radio flux; and \\
(b) Discrete ejecta, separating from the nucleus superluminally at
$\geq$1.3~c, during steep spectrum flares. 

The major outbursts appear qualitatively different from the 30~min
variations, and seem to have a different trigger.  For {\bf flares},
the onset of radio emission seems to correlate with the {\bf rise} of
2-12~keV X-rays (see Figs.~\ref{fig-1}~\&~\ref{fig-4}). On the other
hand, the AU-scale (30~min) radio/IR oscillations are seen when
deep XTE {\bf dips} of $\le$5000 counts are present.  On the short
timescale of 10's of minutes, isolated (single) events have not been
observed to our knowledge - they are always part of a pulse-train,
accompanied by X-ray dips.  On the other hand, no periodicity or
regularity has been observed in the timing of the large flares, which
are always isolated events. More details of the two phenomena follow.

\subsection{The AU-Scale Jet in the Nucleus \label{sec-auscale}}

Our images are consistent with a conventional model of conical
expanding jet, (\citet{hj88,fal99}, see also Sec.\ref{sec-discussion})
i.e.,  incoherent synchrotron emission in an optically thick region of
size $\sim$10$\lambda_{\rm cm}$~AU,
 inclined at $\sim$66-70$^\circ$ to the line of sight. The brightness
temperature of the jet is  T$_{\rm B} \geq$10$^{9}$K at all
wavelengths. (T$_{\rm B}$=10$^{9}$K at 0.7~cm, where
the minor axis is marginally resolved because the the scattering is
least.  At 2~cm and longer wavelengths, we measure the same brightness 
temperature, but it is a lower limit because 
the scattering size dominates the beam resolution, and the intrinsic
width could be smaller.)

The images clearly identify the AU-scale jet with the {\bf plateau}
state and its quasi-periodic flux variations.  The radio light curve of
the jet is best described as a smoothed response, with time-constant of
$\sim$30~min, to the injection of relativistic plasma, presumably
generated during the X-ray dips, (Figs.~\ref{fig-9} \& ~\ref{fig-10}).
The injection interval is variable, 12-60~mins have been observed on
various occasions. The time delay between the shorter and longer radio
wavelengths is also variable, from 4~min to $\sim$30~min. Possible
causes include changes in the size, expansion rate, or optical depth of
the jet, or variable dynamics of accretion and jet formation. Variable
orientation appears not to be a major contributing factor.

\placefigure{fig-9}

\placefigure{fig-10}    

During a radio flare, one expectation might be that the radio core
would fade as the inner accretion disk was ejected and/or swallowed.
However, we find that the nuclear jet re-establishes itself within 18
hrs of the start of a major outburst, if it disappears at all.

The AU-scale structure is smooth, with little evidence for discrete,
moving ejecta. This can be currently ascribed to the difficulty of
Fourier synthesis imaging of moving, time-variable sources, rather than
to a true lack of relativistic flow. A static structure is unlikely,
given that we see superluminally moving ejecta on larger scales  {\em
during the same event.} A more likely explanation is a steady state
with fast, continuous flow and adiabatic losses which cause the jet to
fade rapidly with distance.  The travel-time along  the length of the
jet ($\sim$10~AU~hr$^{-1}$) at relativistic speed is comparable to the
adiabatic loss timescale ($\sim$30~min) over which new IR/radio emission
fades away after each X-ray injection event. Analogous compact jets may
be found in Cygnus X-1 and Cygnus X-3, as suggested by \cite{fen20}
from the flat radio-millimeter spectra of the cores in these X-ray
binaries.

Good evidence exists that the synchrotron spectrum extends up to 2$\mu$
in the IR \citep{eik98,mir98,fen98}. The IR light curves are quite
similar to the radio, believed to be caused by rapid adiabatic
expansion rather than synchrotron losses, as suggested by \citet{fen98}
and \citet{mir98}.  Maximal internal energy \citep{fal99} implies
expansion at $\sim$~0.6~c. The VLA measured expansion rate
\citep{rod95} is about 0.2~c on large scales.  On AU scales, we see an
elongated structure  with length about four times the width, which is
consistent with relativistic flow along the major axis and lateral
expansion at 0.2~c.

Considering next the position angles of the AU-scale images, no  significant
time variations are seen.  The
AU-scale position angle is stable within $\sim$5$^{\circ}$ for $\sim$2
years.  The 500~AU ejecta have the same P.A. for our two measurements
in 1997 and 1998. The large ejection reported in Mirabel \& Rodr\'\i
guez (1994) was along $\sim$150$^{\circ}$, about 7$^{\circ}$ CCW.  We
measure  a rotation of about 12$\pm$5$^{\circ}$ from the few-AU to
the  500~AU scale, though both seem to be stable in time.  This
rotation of position angle with size scale appears real. We suspect it
could be due to opacity effects in a conical jet, since the high
resolution is obtained at shorter wavelengths - however we do not have
a clear explanation for it.

In summary, a synchrotron jet (see also Discussion) accounts in a
unified, consistent  way for the following phenomena:
\begin{itemize}
\item The flat radio spectrum and high brightness temperature.
\item The elongation of the core along the axis of arcsecond-scale
	superluminal ejecta, ($\sim$155$^{\circ}$). 
\item The peak emission is
	progressively delayed at longer wavelengths, as they emanate
	from further along the expanding jet;
\item There is progressively less variation of the flux at longer
 	 wavelengths, due to convolution over a larger region;
\item The decay time of flux variations is consistent with the  
	travel time of relativistic plasma along the jet.
\end{itemize}

However, we must point out that the ratio of integrated flux density
of  approaching to receding parts  of the continuous jet are only
1.15$\pm$0.04, 1.20$\pm$0.05, and  1.10$\pm$0.08, for the three images
of Fig.~\ref{fig-5}A, B, C respectively. (Assuming we are seeing
both sides of the jet; see the next paragraph for a possible exception).  
Assuming an inclination
$\theta$=70$^{\circ}$, a flat spectral index ($\alpha$=0), and a
continuous jet (k=2), we obtain a mildly relativistic speed of
$\beta_{\rm F}$~=~0.1 for the AU-scale jet, from the flux ratio (see,
e.g., \citet{bod95}, eqn.4, for a stationary jet, with pattern speed
$\beta_{\rm S}$~=~0, and bulk flow at speed $\beta_{\rm F}$):

\begin{equation}
\left(  {\rm S}_{\rm approaching} 
\over   {\rm S}_{\rm receding}  \right)  = 
\left(1 +\beta_{\rm F} ~{\rm cos}\theta \over  1 - \beta_{\rm F} 
~{\rm cos}\theta   \right) ^{k-\alpha}
\end{equation} 

This result of 0.1c from the jet/counter-jet asymmetry is puzzling,
in light of the expected relativistic flow in the jet model.
We offer three possible explanations, none completely 
satisfactory: 

\paragraph{Slow Mini-Jet?}
 The jet/counter-jet ratio on AU scales is really due to slow jet 
velocity of 0.1c. This contradicts
some of the data, and we do not favour it. First,
the measured speed of 0.9c at 500~AU
implies acceleration of ejecta from 0.1c to 0.9c between 20~AU
and 500~AU,  very far from 
the black hole ($>$10$^8$R$_{\rm s})$. This seems very unlikely. 
Secondly, as discussed  above,
the expansion speed of the synchrotron cloud must be fast, in at least 
one dimension, to account for the rapid fading of emission from adiabatic loss.
(synchrotron radiative decay would take years, not 30~min as observed). 
The size and elongation of the jet, $\sim$20~AU, implies the speed
is $\sim$c, in at least one direction. For jet speed of 0.1~c, the
isotropic expansion would dominate, so the AU-scale images would 
be circular, and endure for over 10~hrs.

\paragraph{Fast Mini-Jet?}
On the other hand, \footnote{We acknowledge discussion with Peter
Goldreich.} the brightness temperature of
an optically thick synchrotron source should be about 10$^{11}$~K, 
(predicted by the Falcke model as well) for a bulk Doppler factor
$\sim$1, whereas we measure 10$^{9}$~K at 0.7~cm. 
If we assume a fast mini-jet, with bulk $\beta_{\rm F} \geq$0.99,
then, because of the large inclination of 70$^{\circ}$, 
the (approaching) doppler factor is $\leq$0.2 and  the
observed brightness temperature is reduced.
This argues for a highly relativistic mini-jet on scales of few
AU, that slows down as it reaches 1000 AU. Unfortunately, there 
is no observational evidence for deceleration, such as brightening due 
to hitting a density enhancement, at that radius. The jet/counter jet
ratio would require explanation as in the next paragraph.

\paragraph{Hidden Counter-Jet?} 
We can postulate that the (receding) counter-jet is hidden by 
free-free absorption in the inclined disk, and only the
approaching mini-jet is observed, with nearly symmetric brightness
profile. This absorption 
 geometry has been clearly 
demonstrated in 3C84, \citep{wal00}. In GRS~1915+105, the 
absorption must occur
within 1000~AU of the core, because the counter-ejection is
seen outside that radius, \citep{fen99,mir94,rod99} 
We detect no counter-ejection at $<$500~AU in 1998 May 02 
($\tau >$1.5 at 2~cm), and
on 1997 October 31 there is a weak counter-ejection at a flux reduced from 
the expected beaming model, \citep{mir94} indicating $\tau \sim$0.5 at 
3.6~cm, (sec.3.1 and 3.2). Different opacity for each ejection could be
due to variable ionizing illumination from the disk, or variable
mass-loss, for both of which ample evidence exists.
Thus free-free opacity hiding the counter-jet on AU-scales, and hiding 
the counter-ejecta 
within a few 100~AU of the core, is consistent with the images. 
Assuming $\tau_{\rm free-free} \sim$1,  L$\sim$100~AU, we derive an
electron density n$_{\rm e}$=5~10$^{5}$~cm$^{-3}$, emission measure
EM$\sim$2~10$^8$~cm$^{-6}$~pc,   and 
column~density~$\sim$10$^{21}$~cm$^{-2}$, for an electron temperature 
T$_e$=10$^4$~K.

Furthermore, we might explain both the enhanced scattering
(Section-\ref{sec-scat}) and the
free-free opacity as signatures of the ionized envelope 
detected by \citet{mart20}.
However, the scattering is measured against the core, 
and the absorption is invoked against the counter jet, \em not 
\rm the core, so special geometry is needed. We leave this as a speculation.
The test would be to actually
observe the counter jet and measure its spectrum which should be strongly
rising with frequency, since the opacity is exponentially less at shorter
wavelength, \citep{wal00}.  Sensitive, technically difficult 
observations would be required at 0.7~cm or 0.3~cm.

\subsection{The Ejecta at $\sim$500~AU \label{sec-largescale}}

Images during flares show rapidly decaying superluminal ejecta.
Snapshots taken a few hours apart allow us to measure, for the
approaching ejection, a velocity of 1.28$\pm$0.07~c~$({\rm D}/10~{\rm
kpc})$ at 500 to 600~AU away from the nucleus,  consistent with the
velocity of \citet{fen99} at 5000~AU  for  the same event.  This 
implies ballistic motion from
500~AU to 5000~AU.  Phase-referencing clearly distinguishes the
stationary core from the moving ejecta, unlike previous observations.
Next, back-extrapolation of the position of the ejecta (over 2-3
days), using the velocity measured over a few hours, gives a position
coincident with the core at the  start time of the flares. Thus, on
average, this is consistent with ballistic motion from the few-AU scale
outwards.  The lateral expansion rate of the ejecta is about 0.14-0.2c,
and is not well constrained in the VLBA images due to the lack of
sensitivity to extended structure. The value of 0.2~c was measured in
the VLA images on larger scales, \citep{rod99}.

\section{Discussion \label{sec-discussion}}

In this section we discuss a possible thermal origin for the emission,
and decide in favor of a synchrotron jet. We also find good consistency
between the synchrotron model of \citet{fal99} and the disk instability
model of \citet{bel97a,bel97b}  for the source of the relativistic
plasma in the baby jets. The large ejecta require $\sim$10 times more
mass, and may be qualitatively different from the AU-scale jets as well.

\subsection{Thermal Jet}

We rule out a thermal origin for the emission in favor of a synchrotron
jet by the following argument.  If we assume that the radio emission is
due to free-free (thermal) radiation from gas with an electron
temperature of $\sim$10$^{10}$K, and follow the formulation of Reynolds
(1986) for the mass loss rate in a thermal jet, we roughly estimate

\begin{equation} 
\left( \dot{\rm M} \over {\rm M}_{\odot}~{\rm yr}^{-1} \right) 
= 4 \times 10^{-4} 
\left( {\rm v} \over 10^4~{\rm km~s}^{-1} \right) 
\end{equation} 
where v is the velocity of the outflowing gas. If this velocity were of
the order of the speed of sound in a gas of 10$^{10}$K,
v=10$^4$~km~s$^{-1}$, then a mass loss rate of
$\sim$4x10$^{-4}$M$_{\odot}$~yr$^{-1}$, a mechanical power of
3$\times$10$^6~{\rm L}_{\odot}$, and an X-ray luminosity of the jet of
$\sim10^7~{\rm L}_{\odot}$ are derived. These powers are similar to the
Eddington luminosity of a $\sim 100~{\rm M}_{\odot}$ black hole, and
appear excessive for what is known of GRS~1915+105.

Finally, the QPO's observed in the radio and X-rays suggest that the
material is travelling and expanding at relativistic speeds.  This
would make the estimated mechanical power and X-ray luminosity of the
jet increase by a further large factor. We hence favour the synchrotron
emission mechanism for the AU-scale jet.

\subsection{Conical Synchrotron Jet}

A jet model with minimum free parameters is discussed in
\citet{fal99}.
 Considering the coupled disk-jet system, subject to the assumption of
total equipartition of energy between disk, jet kinetic energy, jet
internal energy (thermal energy of proton/electron plasma) and magnetic
field, expanding freely into vacuum, they derive a relativistic jet of
terminal $\beta ~ \sim$0.96. (A higher velocity would require
acceleration by a mechanism other than pressure gradient.) The only
free parameter is the measured disk X-ray luminosity.  In the case of
GRS~1915+105, with X-ray luminosity of 10$^{39}$~erg~s$^{-1}$, (0.3 to
3 times this have been observed) the following properties are expected
for the jet at 2~cm.  The mechanical power of the jet is of the order of
Q$_{\rm jet}$=10$^{39}$~erg~s$^{-1}$, which indicates a mass accretion
rate of 10$^{19}$~g~s$^{-1}$, assuming a conversion efficiency of 10\%
from gravitational energy. Note that this model assumes a maximally
efficient radio jet for a given disk luminosity, i.e. a radio-loud
jet.  This  implies, as pointed out in \citet{fal99}, that
10$^{39}$~erg~s$^{-1}$ is a {\bf lower limit} to the mechanical power
of the jet. The model then predicts a (steady state) flux density of
about 20~mJy, and jet length of about 0.4~mas, with expected wavelength
dependence of jet size
 $\Theta~\propto \nu^{-1}$, and spectrum S$_{\nu}~\propto \nu^{0.2}$.  Other
derived parameters are a magnetic field  B of 0.4~gauss at 10~AU from
the black hole, and an electron Lorentz factor $\gamma _{\rm e}$ in the
range of 200-800.

Thus, in the steady state, the model yields roughly the correct size,
spectrum, and flux of the  AU-scale nuclear jet. Time variable jet
length and flux density, can perhaps be accomodated as changes in mass
accretion rate. Allowing for the complex time-variability in this
system, we judge the agreement with our observations to be
satisfactory.  Next, given that the jet is already maximally radio
loud, it seems to us necessary to invoke an additional mechanism, to
explain the transient release of excess synchrotron emission in the
flares, with spectrum S$_{\nu}~\propto \nu^{-0.6}$. The dependence of flux
density on orientation, (as in AGN unification schemes), does not apply
here since we observe the axis position angle does not change
sufficiently from small to large scales.

\subsection{Disk Instability as the Source of Relativistic Plasma}

GRS~1915+105 shows extremely complex behavior in X-rays, see e.g.
Fig.~\ref{fig-10}, and \citet{mar99}. The  X-ray variability is
explained by \citep{bel97a}, and \citep{bel97b}, in terms of the
evacuation and refilling of the inner hot accretion disk due to
thermal-viscous instability. They attempt to reduce the complexity of
the light curve to essentially one parameter, the radius of the missing
inner disk.  The light curve, and hardness ratio variations, are
decomposed into a superposition of an inner, 2.2~keV disk component, of
variable radius ($\sim$20-100~km), along with an outer constant disk of
radius over 300~km and temperature 0.5~keV. The instability
(emptying/refilling of the disk) occurs in the region where radiation
pressure dominates the gas pressure. The time required to refill the
disk with X-ray emitting gas is set by the viscous/thermal  timescale.
Assuming a viscocity parameter $\alpha \sim$0.01, they estimate this
timescale as

\begin{equation} 
t_{\rm viscous} = 3 \alpha^{-1} 
	\left( {\rm M}_{\rm BH} \over 10 {\rm ~M_{\odot}}  \right)
	\left( {\rm R} \over 10^{7}{\rm cm}  \right) ^{7/2}
	\left( {\rm \dot{M}} \over 10^{18}{\rm ~g~s}^{-1}  \right)^{-2} ~
 {\rm seconds}
\end{equation}       

For the 30~min dips in the X-rays (the timescale of the radio
oscillations) the radius of the missing disk is of the order of 180~km.
From spectral fitting to the disk/black body, they also derive a mass
accretion rate of $\sim$10$^{18}$ to $\sim$10$^{20}$g~s$^{-1}$,
depending on whether the black hole is maximally rotating or not.  The
highest accretion rates are  inversely correlated with the X-ray
luminosity, which they take as evidence that the majority of the mass
may be eliminated from view by an ADAF-type flow into the black hole.

From our AU-scale images, we find good 
agreement between the mass of synchrotron-emitting plasma in the AU-scale
`mini-jet',  and the mass loss from the missing radius of the inner
accretion disk. The timescales and radio/soft X-ray correlations are
consistent with the disk instability model. In contrast, the large 
flares  eject an order of magnitude more mass,  and the recovery 
timescale of $\sim$18~hrs indicates a radius of $\sim$600~km for 
the missing disk. This is so large as to lie outside the radiation-
pressure dominated region, so the instability (oscillation) would not work.
Indeed, the large-scale events never repeat, unlike the AU-scale 
30~min cycles.  Once again, we take this as evidence of a qualitative 
difference between the AU-scale emission and the extreme
ejections during flares. Thus the Belloni model works for mini-jets, 
but not for large flares.

\section{Conclusions}

From VLBA images of GRS~1915+105 in 1997 and 1998 we conclude the
following:

\begin{enumerate}

\item  The core is always resolved as a compact collimated
jet of $\sim$10$\lambda$$_{\rm cm}$~AU, that has a peak brightness
temperature  T$_{\rm B}\geq$10$^9$K.

\item  The compact jet is observed in the {\bf plateau} state that can
last from days up to several months, which consists of a steady flat
spectrum at radio wavelengths, persistently bright hard X-ray
(20-100~keV) flux, and a faint flux in the softer X-rays (2-12~keV)
(see Figs.~\ref{fig-1} \& \ref{fig-2} and
     Figs.~\ref{fig-4} \& \ref{fig-5}).

\item  The compact jet is observed a few hours before and a few hours
after major flare/ejection events (Figs.~\ref{fig-2} \& \ref{fig-5}).
The nuclear jet is re-established in $\leq$18 hours after a major
outburst. In minor outbursts the jet flux varies in $\sim$30~min
(Figs.~\ref{fig-9} \& \ref{fig-10}), with the same period seen in the
infrared and in the 2-12~KeV X-rays, (e.g.  Mirabel et al. 1998; Fender
et al. 1998; Eikenberry et al. 1998).

\item  Although we could not measure the proper motion of matter in the
quasi-steady compact jet, its length of $\sim$20~AU (e.g.
Fig.~\ref{fig-5}D \& E) is comparable to the distance that would
be traversed at relativistic speed during the
time-scale of the oscillations ($\sim$30~min).

\item  The compact jet is optically thick synchrotron emission. This is
suggested by the high degree of collimation, the high brightness
temperature, the flat radio spectrum, and the inferred relativistic
bulk motions. The mass outflow assuming a synchrotron jet is
$\geq$10$^{-8}$M$_{\odot}$~yr$^{-1}$, consistent with the accretion
rate estimated  from the X-ray luminosity of the black hole binary.
Thermal processes alone cannot account for the radio emission because
they are relatively inefficient radiative mechanisms that would require
an accretion rate {\bf greater} than 10$^{-5}$~M$_{\odot}$~yr$^{-1}$,
which is unreasonably large.

\item  The large mass ejection events with measured superluminal
velocities take place at the time of abrupt changes in the X-ray state
of the source.  Within $\pm$5$^{\circ}$ the same position angles were
observed for major ejections in 1994-1996 \citep{rod99}, and 1997-1998
(this paper).  The change from radio-quiet to
radio-loud states cannot be caused by change in  orientation, 
since the position angle of the ejecta does not change
sufficiently from small to large scales. The axis of the ejecta at
large scales ($\geq$500~AU) appear to be rotated clockwise by
$\le$12$^{\circ}$ relative to the jet axis at $\leq$100~AU, both
measured at the same time.  The flux variations must be intrinsic.

\item  Time-lapse images allow us to detect the motions of the
large-scale ejecta within a few hours, with apparent speeds of
$\geq$1.3~c, consistent with previous observations (Mirabel \& Rodr\'\i
guez, 1999).  Knowing the time for the onset of the radio outburst, it
is inferred that the acceleration to terminal speeds must take place
within 500~AU of the black hole.

\item  By astrometry over two years relative to an extragalactic
reference, we locate the black hole binary within $\sim$1.5~mas, and
follow its secular parallax due to Galactic rotation. The black hole is
stationary on the sky to $\le$100~km~s$^{-1}$,  once the secular
parallax is accounted for.

\item  Some accreting black holes of stellar mass (e.g. Cyg~X-1,
1E~1740-2942, GRS~1758-258, GX~339-4) and the supermassive black hole
Sgr~A$^*$, show compact radio cores with a steady flat-spectrum plateau
state, in common with the core of GRS~1915+105. It is possible that
high resolution images of the other systems will also reveal compact
jets. So far GRS~1915+105 is the only galactic source where both,
AU-sized jets and large-scale superluminal jets have been unambigously
observed.

\end{enumerate}

\acknowledgements

We thank the ASM/RXTE team for providing us with quick-look results.
LFR acknowledges support from CONACyT, M\'exico. IFM acknowledges
support from CONICET, Argentina.

\clearpage \newpage

\figcaption[fig-1.eps]{X-ray and Radio data for 1997 October 14 to
November 08 (MJD=50735 to 50760).\protect\\[2mm]
Top:~[red]~RXTE All-Sky Monitor counts, 2-12~keV; quick-look results
provided by the ASM/RXTE team; ~~~[blue]~BATSE~fluxes, 20-100~keV from
occultation measurements, (photons~cm$^{-2}$~s$^{-1}$, multiplied by
2000 for plotting).   \protect\\[2mm]
Middle:~GBI~radio~flux densities at 13 and 3.6~cm.  \protect\\[2mm] 
Bottom:~GBI~radio~spectral index.  The radio data is typically $\sim$10
scans per day of 5~min each, separated by 30-60~min.  Note the
correlation between the start of the soft X-ray and radio flares, and
the similar decay times of the BATSE and radio fluxes.  The VLBA
observed in the periods indicated by the black V's labelled A, B, \& C,
with the corresponding images in Fig.\ref{fig-2}. The vertical green
line is our estimate of the start of the flare at
MJD=50749.8$\pm$0.1~day.
\label{fig-1}} 

\figcaption[fig-2.eps]{VLBA images during 1997 October, corresponding
to epochs A, B, \& C in Fig.~\ref{fig-1}. Contours are at -5\%, 5, 10,
20, 40, 60, \& 90\% of peak.  \protect\\[2mm]
(A)~2~cm~image with 7.5~AU resolution, during a plateau state, showing
the jet in the nucleus, 6 days before the flare in B \& C;
 peak~=~15.7~mJy~beam$^{-1}$.  \protect\\[2mm] (B)~13~cm~image with
220~AU resolution; peak~=~93.7~mJy~beam$^{-1}$.  \protect\\[2mm]
(C)~3.6~cm~image, with 80~AU resolution; peak~=~19.7~mJy~beam$^{-1}$.
\protect\\[2mm] B and C are simultaneous data, centered at epoch
MJD=50752.02, 53.3$\pm$2.4 hours after the estimated start of the
flare.  The position of {\bf A}  is the same as the central component
of {\bf C}  within 1.5mas.  The black `*' marks  the astrometric
position of the nucleus, here and in
 Fig.~\ref{fig-7} A, item 5.
\label{fig-2}} 

\figcaption[fig-3.eps]{Images at 3.6~cm, of the same data as in
Fig.\ref{fig-2}C, epoch 1997 Oct 31. Contours are at -4\%, 4, 8, 16,
32, 64, 96\% of peak.  The SE component has a proper motion of
2.3$\pm$0.2~mas during 2.5~hrs, or 0.92$\pm$0.08~mas~hr$^{-1}$. The
mean position of the SE component implies a motion of 47.5$\pm$0.3~mas
in 53.3$\pm$2.4~hrs since the start of the flare, or
0.90$\pm$0.05~mas~hr$^{-1}$.
\label{fig-3}}

\figcaption[fig-4.eps]{X-ray and radio data for 1998 April 07 to May 12
(MJD=50910 to 50945).  Details are as in Fig.~\ref{fig-1}. Images from
VLBA data, indicated by the black V's labelled D, E, \& F, are shown in
Fig.~\ref{fig-5}. The vertical green line is our estimate of the start
of the flare at MJD=50932.7$\pm0.3$~days.
\label{fig-4}} 

\figcaption[fig-5.eps]{VLBA images during 1998 April-May, corresponding
to epochs D, E, and F in Fig.~\ref{fig-4}.  Contours are at -2\%, 2, 4,
8, 16, 32, 64, 96\% of the peak. \protect\\[2mm]
(D)~2~cm~image, 7.5~AU resolution, showing the nuclear jet during a
plateau state, 2~days before the flare on MJD~50916; 
peak~=~48~mJy~beam$^{-1}$.  \protect\\[2mm]
(E)~2cm~image, 7.5~AU resolution; peak~=~64~mJy~beam$^{-1}$.  
\protect\\[2mm]
(F)~0.7cm~image, 2.5~AU resolution; peak~=~66.4~mJy~beam$^{-1}$.  
\protect\\[2mm]
E and F are quasi-simultaneous data (see Table-\ref{table-1}), centered
at MJD~50935.50. The position of {\bf E} is  identical to {\bf D}
within 1.0~mas. The black `+' marks the astrometric position,  here and
in Fig.~\ref{fig-7}~A, item 7.
\label{fig-5}} 

\figcaption[fig-6.eps]{Lower resolution images at 2~cm, of the same data
as in Fig.\ref{fig-5}, epoch 1998 May 02 at 75~AU resolution.  Contours
are at -2\%, 2, 3, 4, 6, 8, 16, 32, 64, 96\% of peak. The blue and red
contours show time-resolved images 4.5~hrs apart. The SE component has
a proper motion of 4.2$\pm$0.3mas, or 0.93$\pm$0.07~mas~hr$^{-1}$. The
mean position of the SE component implies a motion of 57.5$\pm$0.5~mas
in 67$\pm$7~hrs since the start of the flare, or
0.87$\pm$0.10~mas~hr$^{-1}$.
\label{fig-6}} 

\clearpage \newpage

\figcaption[fig-7.eps]{A:~Astrometric positions from 1996 May, to 1998
May, showing that the motion of GRS1915+105 lies mainly in the galactic
plane. Key:  \protect\\[2mm]
[1]=1996 May 23, 3.6~cm.          [2]=1996 Aug 04, 3.6~cm.
[3]=1997 May 15, 3.6~cm, rise.    [4]=1997 May 15, 3.6~cm, set.
[5]=1997 Oct 23, 2~cm, pre-flare. [6]=1997 Oct 31, 3.6~cm, during flare. 
[7]=1998 Apr 11, 2~cm, pre-flare. [8]=1998 May 02, 2~cm, during flare. 
The sloping line is parallel to the Galactic plane.  \protect\\[2mm]
B:~Expected~secular~parallax from galactic rotation vs. distance, in
the direction of GRS1915+105, assuming 220~km~s$^{-1}$ rotation, (solid
line).  If the source were at 12~kpc, and stationary with respect to
the galactic surroundings, 5.3~mas~yr$^{-1}$ is expected, compared to
the 5.8$\pm$1.5~mas~yr$^{-1}$ measured. The effect of a $\pm$10\%
($\pm$22~km~s$^{-1}$) variation in rotation velocity is shown as the
dashed lines.
\label{fig-7}} 

\figcaption[fig-8.eps]{Deconvolved core size (Gaussian FWHM) vs.
wavelength. Galactic electron scattering is fitted to the intrinsically
unresolved minor axis, (lower line) $\Theta_{\rm min}=(0.15\lambda_{\rm
cm}^{2}$)mas.  \protect\\[2mm]
An intrinsically elongated source, plus scattering, is independently
fit to the major axis,  (upper line), $\Theta_{\rm Maj}=(1.0\lambda_{\rm cm}+0.14\lambda_{\rm cm}^{2}$)mas. \protect\\[2mm] 
The solid lines are fitted to data from a single day (1996 Aug 01) when
the flux from the core was steady. Dashed lines extend to newer data
(1998 May 02) at 2~cm and 0.7~cm. The Y-axis at right shows the linear size
at an assumed distance of 12~kpc. The jet length is variable by up to a
factor of $\sim$2 at different epochs.
\label{fig-8}} 

\clearpage \newpage

\figcaption[fig-9.eps]{Radio flux density variations at three
wavelengths, monitored by the VLA during the plateau state on
1997~May~15, 06-16~UTC. Inset is the VLBA at 2~cm, from observations
during the gap in the VLA coverage from 10-12~UTC.
\label{fig-9}}

\figcaption[fig-10.eps]{Top: XTE 2-12keV counts on 1997 Sep 05,
06-10~UTC. Note the dips every $\sim$12~min.  \protect\\[2mm]
Bottom:~VLA~flux density at 3.6~cm, with the same periodicity (within
errors) as the X-ray dips.  Fig.~\ref{fig-9} and Fig.~\ref{fig-10}
taken together, support the idea that the radio oscillations arise in a
synchrotron jet of (Gaussian FWHM) size $\sim$10$\lambda$~AU, which is
fed by plasma injected during the dips in the soft X-ray flux.
\label{fig-10}}

\clearpage \newpage

\begin{table}
\begin{tabular}[]{lllllllll}
\tableline               
\tableline               

 & ~~~MJD & ~~~~~~~~~~UT & Fig. & ~$\lambda$ & Bit rate 
& rms, M. & rms, T. & beam \\ 

 &   &  &   & (cm)    & (Mb~s$^{-1}$) 
& ($\mu$Jy) & ($\mu$Jy)  & (mas)  \\

\tableline               
A & 50744 & 97 Oct 23 22:15-03:23 & \ref{fig-2} A & 2.0 & 256
& ~210 & ~85 & 0.62 \\

B & 50751 & 97 Oct 30 23:00-04:08 & \ref{fig-2} B & 13  & 128$^B$
& ~180 & ~75 & 20. \\

C & 50751 & 97 Oct 30 23:00-04:08 & \ref{fig-2} C & 3.6 & 128$^C$
& ~200 & ~75 & 7.0 \\

D & 50914 & 98 Apr 11 09:00-14:08 & \ref{fig-5} D & 2.0 & 256
& ~230 & ~85 & 0.62 \\

E & 50935 & 98 May 02 07:35-15:55 & \ref{fig-5} E & 2.0 & 128$^E$
& ~260 & 120  & 0.62 \\

F & 50935 & 98 May 02 08:26-16:22 & \ref{fig-5} F & 0.7 & 256$^F$
& ~510 & 300  & 0.21 \\
\tableline               
\end{tabular} 

\vspace{5mm}

B, \&  C in Fig. \ref{fig-2}  were  simultaneous observations,
	128~Mb/s was recorded at each of 3.6~cm and 13~cm. \\
E, \&  F in Fig. \ref{fig-5}  were quasi-simultaneous observations,
	the sequence (128~Mb/s for 44~min at 2~cm, 5~min dead time, 
	then 256~Mb/s for 22~min at 7~mm) was repeated 7 times.

\vspace{2mm}

\caption{VLBA observations. The columns are: Observation Epoch; 
Figure reference; Wavelength; Bit
rate, megabits per second; rms noise in image, microJansky per beam
area, measured; RMS noise, theoretical; and Synthesized Beam FWHM, mas.}
\label{table-1}    
\end{table}

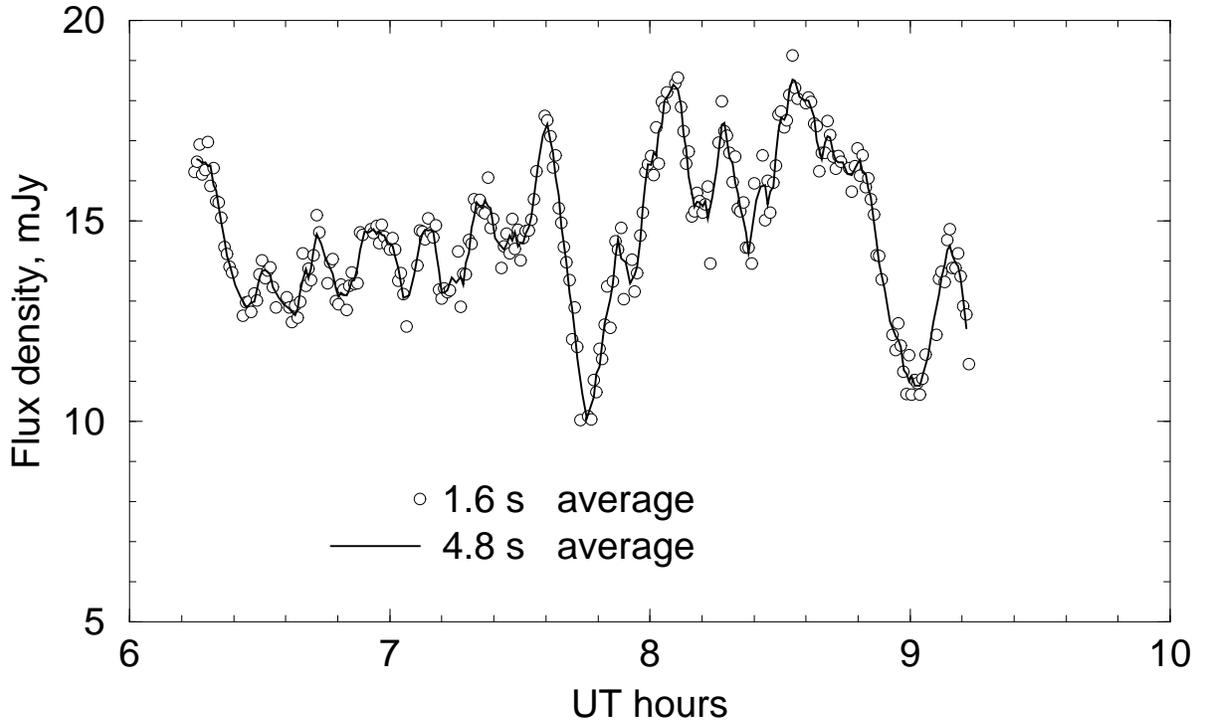
\end{document}